%% file: ramc.tex
\documentclass[sigconf,nonacm]{acmart}

\usepackage{algorithmic}
\usepackage{graphicx}
\usepackage{textcomp}
\usepackage{booktabs}
\usepackage{xcolor}
\usepackage{tabularx}
\usepackage{orcidlink}
\usepackage{listings}
\usepackage{caption}
\usepackage{subcaption}
\usepackage{cleveref}

\definecolor{codegreen}{rgb}{0,0.6,0}
\definecolor{codegray}{rgb}{0.5,0.5,0.5}
\definecolor{codepurple}{rgb}{0.58,0,0.82}
\definecolor{backcolor}{rgb}{0.95,0.95,0.92}

\lstdefinestyle{mylistingstyle}{
    backgroundcolor=\color{backcolor},   
    commentstyle=\color{codegreen},
    keywordstyle=\color{magenta},
    numberstyle=\tiny\color{codegray},
    stringstyle=\color{codepurple},
    basicstyle=\ttfamily\footnotesize,
    breakatwhitespace=true,         
    breaklines=true,                 
    captionpos=b,                    
    keepspaces=true,                 
    numbers=left,                    
    numbersep=5pt,                  
    showspaces=false,                
    showstringspaces=false,
    showtabs=false,                  
    tabsize=2
}

\begin{document}

\newcommand{\wws}[1]{\textcolor{orange}{[WWS: #1]}}
\newcommand{\mgfd}[1]{\textcolor{purple}{[MGFD: #1]}}
\newcommand{\sll}[1]{\textcolor{violet}{[SLL: #1]}}

\title{RAMC: Remote Access Memory Channels over HPE Slingshot}

\author{Whit Schonbein \orcidlink{0000-0003-4955-2984}}
\email{wwschon@csandia.gov}
\orcid{0000-0003-4955-2984}
\author{Matthew G.F. Dosanjh \orcidlink{0000-0001-5141-9176}}
\email{mdosanj@sandia.gov}
\orcid{0000-0001-5141-9176}
\author{Scott Levy \orcidlink{0000-0002-2232-3201}}
\email{sllevy@sandia.gov}
\orcid{0000-0002-2232-3201}
\affiliation{%
  \institution{Sandia National Laboratories}
  \city{Albuquerque}
  \state{New Mexico}
  \country{USA}
}

\renewcommand{\shortauthors}{Schonbein et al.}

\input{abstract}

\ccsdesc[500]{Networks}
\ccsdesc[500]{Networks~Network architectures}
\ccsdesc[500]{Networks~Programming interfaces}
\keywords{Networks, Network APIs, Message Passing Interface, Slingshot, Remote Memory Access}

\maketitle
  
\input{introduction}

\input{motivation}

\input{design}

\input{api}

\input{example}

\input{evaluation}
\input{related}

\input{challenges}

\begin{acks}
Sandia National Laboratories is a multi-mission laboratory managed and operated 
by National Technology \& Engineering Solutions of Sandia, LLC (NTESS), a wholly 
owned subsidiary of Honeywell International Inc., for the U.S. Department of 
Energy’s National Nuclear Security Administration (DOE/NNSA) under contract 
DE-NA0003525. This written work is authored by an employee of NTESS. The 
employee, not NTESS, owns the right, title and interest in and to the written 
work and is responsible for its contents. Any subjective views or opinions that 
might be expressed in the written work do not necessarily represent the views 
of the U.S. Government. The publisher acknowledges that the U.S. Government 
retains a non-exclusive, paid-up, irrevocable, world-wide license to publish 
or reproduce the published form of this written work or allow others to do so, 
for U.S. Government purposes. The DOE will provide public access to results of 
federally sponsored research in accordance with the DOE Public Access Plan. 
SAND2026-20283C
\end{acks}

\bibliographystyle{ACM-Reference-Format}
\bibliography{refs.bib}

\end{document}

%% file: abstract.tex
\begin{abstract}
In this paper, we present Remote Access Memory Channels (RAMC), an 
explicit one-sided communication library designed to leverage 
the capabilities of HPE Cray Slingshot network hardware. Existing one-sided 
communication frameworks, such as MPI RMA and OpenSHMEM, rely 
on monolithic shared memory models that introduce scalability and usability 
challenges. These frameworks often assume symmetric memory regions or require 
blocking collective operations for window creation, which can mismatch user 
communication needs and hinder performance. Implicit models, such as PGAS and 
UPC, aim to simplify programming by treating local and remote memory as a 
unified region but ultimately rely on explicit mechanisms to implement data movement. 
MPI's recently-introduced partitioned communication API offers a 
persistent point-to-point interface but sacrifices the dynamic flexibility 
of RDMA. RAMC is designed to address these limitations. Based on the core 
concept of a persistent uni-directional communication channel, RAMC leverages 
Slingshot's unique memory region counters to enable efficient completion 
notification. Experiments with a RAMC-based heat diffusion code demonstrate RAMC has no difficulty 
scaling to 19.6 thousand processes across 250 nodes, and microbenchmark studies 
across multiple libfabric versions show RAMC can outperform Cray's proprietary 
MPI implementation (e.g., increases in bandwidth ranging from $\sim$100\%-130\% for 
1B-4KiB messages under libfabric 1.15.2, and from $\sim$30\%-45\% under libfabric 2.3.1) while 
identifying additional areas for improvement, such as small message latencies.
\end{abstract}

%% file: introduction.tex
\section{Introduction} \label{sec:introduction}

Traditional point-to-point communication models face growing challenges 
from increasing intra-node parallelism incurred by higher CPU core 
counts and the use of accelerators such as GPUs.
These models, exemplified by MPI's message-passing paradigm, rely on complex branching code paths for message processing and destination negotiation. Such complexity can lead to performance bottlenecks, particularly in highly multi-threaded environments and on processors less suited to branching, such as GPUs. 
To address these limitations, networks like HPE Cray's Slingshot offer advanced 
support for one-sided remote direct memory access (RDMA) communication, enabling 
new possibilities for efficient and scalable communication.

Existing one-sided communication frameworks, such as MPI RMA and OpenSHMEM, rely 
on monolithic shared memory models that introduce scalability and usability 
challenges. These frameworks often assume symmetric memory regions or require 
blocking collective operations for window creation or transitions between epochs, which can mismatch user 
communication needs and hinder performance. Implicit models, such as PGAS and 
UPC, aim to simplify programming by treating local and remote memory as a 
unified region but ultimately rely on explicit mechanisms at some level. 
MPI's recently-introduced Partitioned Communication API offers a 
persistent point-to-point interface but sacrifices the dynamic flexibility 
of RDMA. These limitations motivate the need 
for a new approach that fully exploits modern network hardware capabilities.

In this paper, we introduce Remote Access Memory Channels (RAMC), a novel 
explicit one-sided communication library designed to leverage the flexibility of 
HPE Cray's Slingshot hardware. Based on a core building block of a persistent, uni-directional channel between initiator and target, RAMC enables dynamic 
and flexible communication patterns between processes. RAMC leverages 
Slingshot's unique memory region counters to enable lightweight RMA completion for 
efficient synchronization, addressing key limitations of existing 
one-sided communication frameworks. In general, RAMC provides users with a lightweight, adaptable interface that 
simplifies the transition from traditional MPI point-to-point communication 
while unlocking the full potential of RDMA.

To evaluate RAMC, we conduct correctness tests, scalability experiments 
using a heat diffusion code, and performance benchmarks under two different 
versions of libfabric. 
The scalability experiments show that RAMC has no difficulty 
scaling to 19.6 thousand processes across 250 nodes. Microbenchmark studies 
across multiple libfabric versions show RAMC can outperform Cray's proprietary 
MPI implementation (e.g., increases in bandwidth ranging from $\sim$100\%-130\% for 
1B-4KiB messages under libfabric 1.15.2, and from $\sim$30\%-45\% under libfabric 2.3.1) while 
identifying additional areas for improvement, such as small message latencies.

The contributions of this paper are as follows: 
\begin{itemize}
\item Development of RAMC, a novel one-sided communication library leveraging advanced Slingshot features.
\item Comprehensive exploration of RAMC's design and implementation, including key architectural decisions.
\item Evaluation of RAMC's functionality, scalability, and performance.
\end{itemize}

%% file: motivation.tex
\section{Motivation}
\label{sec:motivation}

\begin{figure}
    \centering
    \begin{subfigure}{0.5\textwidth}
    \centering
        \includegraphics[width=0.5\textwidth]{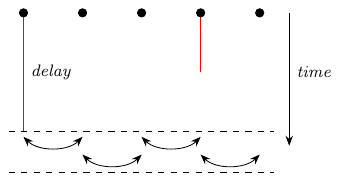}
        \caption{Traditional RMA}
        \label{fig:no-earlybird}
    \end{subfigure}
    \begin{subfigure}{0.5\textwidth}
    \centering
        \includegraphics[width=0.5\textwidth]{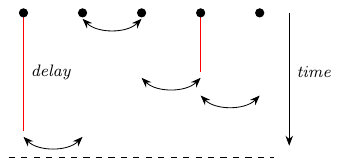}
        \caption{Earlybird}
        \label{fig:earlybird}
    \end{subfigure}
    \caption{A simple `three-point' stencil illustrating how relaxed synchronization may 
    accommodate process delay and create opportunities for communication/computation overlap in comparison 
    to traditional approaches. Dashed lines indicate fences.}
    \label{fig:earlybird-example}
\end{figure}

Interprocess communication based on RDMA is an attractive prospect 
for a variety of reasons, including but not limited to avoiding target OS involvement in data movement 
operations, allowing buffers to be repeatedly reused once registered, and flexibility of data placement via 
offsets and ranges. For example, many modeling and simulation applications exhibit highly regular communication patterns 
(e.g., halo exchanges and collectives), repeatedly interacting with the same peers and exchanging similar data 
volumes into consistent regions of application memory. The setup and coordination of these patterns can be 
expensive, such as in alltoall and allreduce collectives. 
This highlights the need 
for persistence: A reusable communication interface where the overhead of remote access setup is incurred once, and 
the resulting data movement can be efficiently invoked many times. A natural fit for this is one-sided 
communication leveraging the performance of modern RDMA networks. 

However, contemporary user-level communication libraries (such as the Message Passing Interface (MPI)) can 
impose constraints on how RMA is expected to be used, making it difficult to deploy in desired ways, and 
potentially impacting performance. For example, Levy et al.~\cite{levy_leveraging_2024} propose a SmartNIC-based data movement service: During setup, a process on 
a SmartNIC is given access to certain regions of host memory, and when subsequently signaled by the host 
application process, pulls data from those regions for purposes of checkpointing or inserting into a 
distributed database. This seems like a prime candidate for RDMA communication. However, setting up such a 
communication channel is not straightforward. For instance, \texttt{MPI\_Win\_create} is a collective operation within an MPI communicator, 
where each process makes a buffer 
available to every other process in the communicator for 
subsequent RMA operations~\cite{mpi30}. On one hand, because exposing buffers for RDMA operations is \emph{monolithic} 
(in the sense every process in a communicator participates), creating a simple point-to-point relation requires 
down-selecting to a group of size two and creating a new communicator object. 
On the other, even when this is done, the presupposed \emph{mutual involvement}  
of subsequent RMA operations results in a superfluous buffer (on the SmartNIC) that 
may not be used. 

This example motivates an approach to RMA communication that is conceptually closer to traditional 
point-to-point communication than those that treat windows as group-level abstractions (such 
as MPI's RMA windows or OpenSHMEM's symmetric heap). For this alternative approach, the fundamental 
building block of RMA communication should simply be a relation between an initiator (who issues RMA operations) and a 
target (the specified buffer RMA operations act on).

This alternative facilitates potential performance optimizations in group 
communication (such as halo exchanges and collectives) that may be difficult (but not impossible) to achieve under current 
APIs.~\Cref{fig:earlybird-example} shows an example of a three-point stencil, where each process (dot) 
exchanges data with its neighbors to the east and west. A traditional MPI RMA approach (\Cref{fig:no-earlybird}) fences to 
guarantee all prior communication operations have completed for all processes (a collective operation) 
and then engages in data exchange before fencing again. In this contrived example, the first and fourth processes are delayed, 
and all data exchanges are likewise delayed until the first fence completes (dashed line). 

In contrast, if synchronization is not monolithic, and involves only checking that the intended target of 
an RMA operation is ready to handle that request, data exchange can occur in an `earlybird' fashion: While 
some processes may be delayed, others can proceed with their data exchanges as they become ready, as illustrated 
in~\Cref{fig:earlybird}. The result is twofold: First, because data exchanges can occur while other processes are 
delayed, there is less communication to perform when the last process is ready, so the overall group operation 
may complete earlier than otherwise~\cite{marts_cmb_2024}. Second, because processes may receive their data 
while others are still in a communication phase, there is the possibility of doing work with that data, overlapping 
computation and communication.
By reducing the need for global synchronization,  this approach reduces the sensitivity of the composed application to performance perturbation 
and variability, \emph{see} Ferreira et al.~\cite{ferreira:2008:characterizing}.  Although inter-process synchronization may still exist in the form of chains of point-to-point communication operations, the resulting dependencies from these operations impose less strict synchronization, 
\emph{see} Levy et al.~\cite{levy:2016:leveraging}, and allow for communication delays to be 
absorbed rather than being compounded by global synchronization operations.

Finally, we note that modern interconnects such as Cray HPE's Slingshot provide increasingly capable 
one-sided communication support, including independent progress (once data is received by the NIC, it 
will eventually appear in application memory without host involvement) and hardware mechanisms for tracking 
the completion of RDMA operations despite the host being oblivious (e.g., Slingshot's completion counters). 
In contrast, because they are largely organized around monolithic, collectively created windows and group-centric 
setup and synchronization semantics, current one-sided programming interfaces (including MPI RMA 
and OpenSHMEM) make it difficult for users to access the full one-sided communication flexibility 
and performance that the hardware supports.

%% file: design.tex
\section{Design}
\label{sec:design}

In this section we describe the overarching design goals guiding the development 
of RAMC, as well as its core concepts and functionality.

\subsection{Design Goals}
\label{sec:design-goals}

The first goal of RAMC is to avoid monolithic RMA window creation: Consistent with 
a point-to-point communication model, target windows should be able to be exposed to individual initiating processes, and 
without requiring initiating processes reciprocate with an RMA window of their own. Moreover, the window creation process should 
be \emph{non-blocking} to avoid the possibility of deadlock when creating multiple target windows. This in turn implies 
window creation should utilize creation criteria other than call (program) order, such as MPI point-to-point's tag matching 
semantics.

Second, relations between initiators and targets should be genuinely persistent so as to underwrite the repeated use of 
target buffers by fixed or dynamic communication patterns. This is in contrast to, for example, MPI's `persistent communication' API which
caches communication parameters but still requires a new tag match and destination negotiation on each activation and does not establish a persistent 
remote-access relationship. A genuinely persistent building block can lead to more efficient implementations of higher-level API 
abstractions, such as MPI's partitioned communication~\cite{dosanjh2021implementation}. 

Third, RAMC should leverage hardware capabilities to provide efficient RMA with notification~\cite{belli_notified_2015}. Because RDMA bypasses the host OS 
when accessing target memory, host applications are oblivious to the fact an RMA operation has occurred. The traditional way to 
address this is to build a notification mechanism by following the write of data with a second write to a notification buffer, 
relying on underlying ordering guarantees. Contemporary networks may provide mechanisms for avoiding this additional step, e.g., by allowing RMA operations to generate 
events on a target's completion queue or through NIC-based hardware counters.

Finally, RAMC should embrace relaxed synchronization, i.e., targets and initiators can be synchronized pair-wise (initiator/target), even in 
cases where many processes are participating in a global operation such as a collective. If needed, larger scale synchronization can be 
constructed using the more fundamental building block of a point-to-point RMA-based channel.

\subsection{Core Concepts}
\label{sec:design-concepts}

\subsubsection{Counters}
\label{sec:design-counters}

Consistent with the design principle of leveraging Slingshot hardware capabilities, 
RAMC makes use of hardware counters to track the completion of local and remote 
communication operations. Specifically, RAMC leverages two types of counters, which we refer to as \textit{endpoint} and 
\textit{memory region} counters. The former are associated with libfabric 
endpoints (cf. the libfabric man pages for \texttt{fi\_cntr}), and are configured 
to count local \texttt{FI\_WRITE} and \texttt{FI\_READ} operations. 
When incremented, these counters 
indicate the associated operation has 
completed. In the case of a write, the counter incrementing indicates 
the local buffer is available for reuse -- that is, an \texttt{ACK} has been 
received from the target NIC confirming the data has been received -- and in the 
case of a read, the counter indicates the retrieved data is visible 
to the application.

Building on the Portals network programming API~\cite{portals_43}, Slingshot 
also provides counters that can be associated with specific buffers (cf. the 
libfabric man pages for \texttt{fi\_mr}). These 
\textit{memory region} (MR) counters count either the number of remote operations 
that have occurred on the associated buffer (e.g., the number of remote writes), or the 
number of bytes manipulated by such operations (e.g., the number of bytes written). 
RAMC currently uses memory region counters to count operations. Using counters as a lightweight notification 
mechanism is important because it avoids additional followup communication typically required to 
ensure data visibility to a target. 

\subsubsection{Channels}
\label{sec:design-channels}

\begin{figure}
\centering  
\includegraphics[]{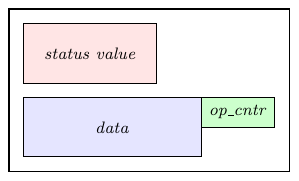}
\caption{Target-side channel data structure.}
\label{fig:channel-target}
\end{figure}  

\begin{figure}
\centering  
\includegraphics[]{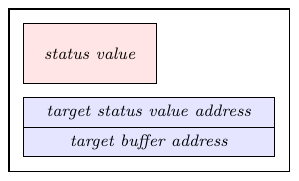}
\caption{Initiator-side channel data structure.}
\label{fig:channel-initiator}
\end{figure}  

Inter-process communication in RAMC is built around the concept of a \textit{channel}, which is a relationship 
between an initiator process and a target window. The target side of a channel is anchored by a structure with two buffers, 
as shown in~\Cref{fig:channel-target}. The \textit{data} buffer is 
the location to which data is written or from which it is read by initiators. Associated with the \textit{data} buffer is 
a memory region counter ($op\_cntr$) used for testing remote operation completion (\texttt{FI\_REMOTE\_WRITE} and \
\texttt{FI\_REMOTE\_READ}). The \textit{status value} 
is an unsigned integer made available to passively coordinate data transfers across the channel by providing information 
regarding the state of the target buffer. At channel creation, the status is initialized to a user-specified value 
($\geq 2$), and the user can manipulate this value using `increment' and `set' commands provided by the RAMC API.
The semantic for this value is determined by the user. For instance, 
examples in this document initialize the status to two, and interpret even status values as indicating the target is in 
an \texttt{OK\_TO\_READ} state, while odd values indicate \texttt{OK\_TO\_WRITE}. Updating (i.e., switching) between states 
is accomplished by incrementing 
the status value by one, which also indicates the target buffer has moved forward a phase in the overall 
communication trajectory of the channels it anchors. In other words, the status value is the means by which 
a user can secure pair-wise initiator/target synchronization.

The initiator side of a channel is anchored by a data structure comprising three pieces of information 
(\Cref{fig:channel-initiator}): (i) Addressing information for a target window's status value, (ii) 
addressing information for that target's data buffer, and (iii) a local status value. The first allows 
an initiator to query (\texttt{fi\_read}) the target's status value, and the second allows an initiator to 
perform puts/writes and gets/reads to and from the target's data buffer. Since Slingshot does not use 
virtual addresses (i.e., it is offset-based), this addressing information comprises the memory keys 
returned by the CXI provider at the time these buffers are registered. The mechanism by which an initiator 
acquires these keys is described below. Mirroring that of the target, the examples in this article use the status 
value held by the initiator to represent (i) the state of the initiator 
relative to that target (`ready-to-send' or `ready-to-receive'), and (ii) its communication phase relative to 
the start of the application run.

\begin{figure}
\centering                                                                      
\includegraphics[width=\linewidth]{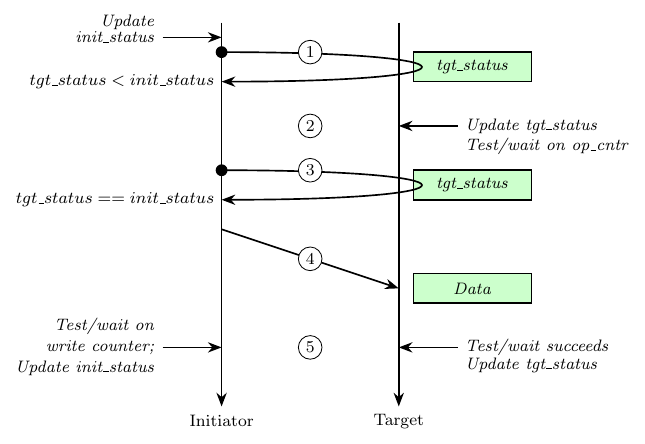}                                   
\caption{A simple write example.}
\label{fig:write-example}
\end{figure}  

An example of how these components come together to coordinate a write from an initiator to a target is 
illustrated in~\Cref{fig:write-example}. In step (1), the initiator is ready to perform a write to the target buffer, so 
it first increments its local status value (from `ready-to-read' to `ready-to-write'), reads the 
target's status, and compares it with its own. The logic guiding the outcome of this comparison is straightforward: 
(i) If $tgt\_status < init\_status$, the the target is behind the initiator, and the initiator should refrain from performing a write. (ii) If $tgt\_status > init\_status$, then the target is ahead of 
the initiator, which may be a catastrophic error (because the target moved on in computation without 
relevant data). Finally, passing these checks guarantees $tgt\_status == init\_status$, in which case the 
write can proceed. This is just one way to make use of the 
target and initiator status values; RAMC is flexible and it allows the 
user to decide how these values are used.

Returning to the example (\Cref{fig:write-example}), step (1) concludes with the initiator determining the 
target is behind in its communication phase, so the initiator is free to do other work. At step (2), the target increments 
its status value to an \texttt{OK\_TO\_WRITE} state. Consequently, when (at step (3)) the initiator returns to 
check the target's status again, it is able to perform the write (step (4)). After 
the write, the initiator may test/wait on the local endpoint counter (configured to 
count write operations) to determine the source buffer may be reused, and updates its status value to 
indicate it has moved past the write phase (step (5)). Similarly, the target's test or wait on the 
operations counter associated with the window will succeed, and the target updates the 
buffer status value again, returning it to an \texttt{OK\_TO\_READ} state.

\subsubsection{The Bulletin Board}
\label{sec:design-bb}

\begin{figure}
\centering                                                                      
\includegraphics[]{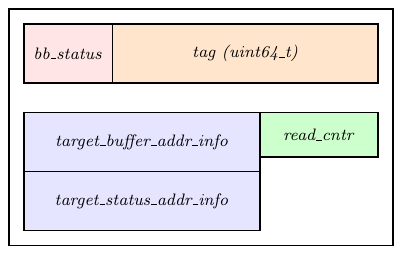}                                   
\caption{The Bulletin Board.}
\label{fig:bb}
\end{figure}

Creating a channel requires that a target register status and data buffers 
(using \texttt{fi\_mr\_[reg,bind,enable]}) to obtain memory keys for those buffers, 
and share that addressing information with initiators who will participate in 
the channel. In RAMC, the exchange of addressing information is accomplished through a 
\textit{bulletin board} (BB) (\Cref{fig:bb}). 

At initialization, a RAMC process allocates space for the BB and shares the addressing 
information for its BB with all other processes (using PMI~\cite{castain_pmix:_2018}). The BB comprises two 
distinct buffers, each of which can be read by any other process. The first 
contains a BB \textit{status} and \textit{tag}, and the second the addressing information for 
a target buffer's status and data buffers. To share addressing information via the BB, 
a target first fills in the status and data buffer addressing information, the tag, 
and then switches the BB status from \textit{inactive} to \textit{active}. The target then 
can test or wait on a memory region counter associated with the BB addressing 
information and configured to count \texttt{FI\_REMOTE\_READ}s until all expected 
initiators have retrieved that information. When this condition is met, the target 
deactivates the BB posting. Similarly, to retrieve addressing information, an 
initiator reads the target's status and tag from its BB. If the status is active, 
the initiator checks to see if the tag matches, and if so, reads the target buffer 
addressing information. If the BB is inactive or the tag does not match, the 
initiator can try again later.

The BB mechanism satisfies the design goal of realizing non-blocking window 
creation, and ensures tag matching occurs only once, during the initial acquisition of 
target addressing information. In its current implementation, targets are limited to 
posting a single BB entry at a time; however, it is trivial to extend the BB to 
accommodate multiple postings or an arbitrary number of postings (via a linked list).

\subsubsection{Discussion}
\label{sec:design-discussion}

It should be clear from this broad summary that RAMC affords significant 
flexibility in how a user can affect inter-process communication. For example, 
the procedure illustrated in~\Cref{fig:write-example} is straightforwardly extended to 
reads by adjusting when the target and initiator status values are updated (so that data transfer 
occurs when the target is in an \texttt{OK\_TO\_READ} state). Second, because 
the semantic for status values is user-defined, RAMC is not limited to 
the two states used by the examples presented here; for example, it is 
simple to define a third state (e.g., \texttt{NOT\_OK}) prohibiting an 
initiator doing \textit{any} remote operation on a target window. Third, 
during a write or read phase, multiple operations may occur, either from the same initiator using the same source 
buffer, the same initiator using multiple source buffers, or from multiple initiators; 
the target simply adjusts the expected value of the operation counter associated with the 
data buffer. Fourth, there are few restrictions on the roles various buffers can play: 
A buffer used as a source for a write or destination for a read can also be a target 
buffer in a RAMC channel, or an initiator can use different buffers for different communication 
operations. Finally, a single initiator can be in different states relative to different targets. It 
may be ready-to-write relative to one target window, and ready-to-read relative to another.

We also explicitly note the current design is \textit{passive target} in the sense 
the initiator reads status information from the target to determine whether the 
target buffer can be read from or written to. In contrast, under 
an \textit{active target} paradigm, an initiator tests or waits for a \textit{clear to send} 
message from a target (e.g., via a write from target to initiator), issuing communication 
operations only once the signal is received. Both approaches have their merits, 
and active targets are intended for future RAMC development.

%% file: api.tex
\section{API}
\label{sec:api}

In this section we provide an overview of the current RAMC API. Functions are 
divided broadly into three categories: Common functions, target-side functions, and 
initiator-side functions.

\subsection{Common Functions}
\label{sec:api-common}

Common functions are those shared by both targets and initiators. These are 
primarily initialization and finalize routines. At startup, 
each RAMC process calls \texttt{ramc\_init()} to select the CXI provider for libfabric, 
create endpoints, allocate NIC resources, share addressing information, set up 
the bulletin board, and so forth. At the time of writing, RAMC attempts to 
select a NIC that is close to each process, although generalizing this mechanism 
is left for future work. The last RAMC call made by each process is 
\texttt{ramc\_finalize()}, which frees any resources not released by other 
calls.

\subsection{Target Functions}
\label{sec:api-target}

\begin{table}
    \centering
    \begin{tabularx}{\columnwidth}{lX}
    \toprule
    \textbf{API} & \textbf{Purpose} \\
    \midrule
    \texttt{create\_window} & Create target window \\
    \texttt{post\_window} & Post window addressing info to BB\\
    \texttt{activate\_bb} & Set BB status to active \\
    \texttt{deactivate\_bb} & Set BB status to inactive \\
    \texttt{await\_bb\_reads} & Wait (blocking) on expected number of BB reads\\
    \texttt{test\_bb\_reads} & Test (non-blocking) for expected number of BB reads\\
    \texttt{await\_win\_ops} & Wait (blocking) on expected number of data buffer communication operations\\
    \texttt{test\_win\_ops} & Test (non-blocking) for expected number of data buffer communication operations\\
    \texttt{increment\_win\_status} & Increment target buffer status\\
    \texttt{set\_win\_status} & Set target buffer status\\
    \texttt{destroy\_window} & Destroy target window\\
    \bottomrule    
    \end{tabularx}
    \vspace{1pt}
    \caption{RAMC Target API. All functions have the prefix \texttt{ramc\_tgt\_} (not shown).}
    \label{tab:target-api}
\end{table}

\sloppypar Target functions are intended to be called by the target of a RAMC channel (\Cref{tab:target-api}). Creating a 
target window (\texttt{ramc\_tgt\_create\_window}) registers the user-provided status and data buffers with the NIC, initializes the target 
status value to a user-defined value, and 
returns a structure that can be used by a target for subsequent 
operations, such as testing or waiting on data buffer communication operations (\texttt{ramc\_tgt\_await\_win\_ops} or 
\texttt{ramc\_tgt\_test\_win\_ops}), updating the target status (\texttt{ramc\_tgt\_increment\_win\_status}, \texttt{ramc\_tgt\_set\_win\_status}), or 
tearing down a channel (\texttt{ramc\_tgt\_destroy\_window}). Bulletin board manipulation functionality -- such as 
posting target addressing information and testing or waiting on an expected number of remote reads of that information -- is 
also provided through the target API. Destroying a window de-registers the data and status buffers. 

\subsection{Initiator Functions}
\label{sec:api-initiator}

\begin{table}
    \centering
    \begin{tabularx}{\columnwidth}{lX}
    \toprule
    \textbf{API} & \textbf{Purpose} \\
    \midrule
    \texttt{check\_bb\_status} & Query BB status of target window. Returns \texttt{RAMC\_SUCCESS} if BB active and tag matches\\
    \texttt{get\_bb\_status} & Retrieve the BB status of target window\\
    \texttt{get\_bb\_posting} & Get addressing information from target BB\\
    \texttt{check\_win\_status} & Returns \texttt{RAMC\_SUCCESS} if target status $==$ initiator status; returns \texttt{RAMC\_AHEAD} if target status $>$ initiator status; returns \texttt{RAMC\_BEHIND} if target status $<$ initiator status\\
    \texttt{get\_win\_status} & Get target window status \\
    \texttt{increment\_status} & Increment initiator status \\
    \texttt{set\_status} & Set initiator status \\
    \bottomrule    
    \end{tabularx}
    \vspace{1pt}
    \caption{RAMC Initiator API. All functions have the prefix \texttt{ramc\_init\_} (not shown).}
    \label{tab:initiator-api}
\end{table}

Initiator functions are called by an initiator in a 
RAMC channel, and include functions for interacting with a target BB 
and window (\Cref{tab:initiator-api}) and communication operations 
(\Cref{tab:communication-api}). 
\texttt{ramc\_init\_check\_bb\_status} returns \texttt{RAMC\_SUCCESS} when 
the status of the target BB is \textit{active} and the tag matches; the `get' 
version simply returns the status and tag. When the BB status is active and the 
tag matches, \texttt{ramc\_init\_get\_bb\_posting} is used to retrieve the 
target status and 
buffer addressing information from the target's BB; this call also initializes 
the initiator's status value to a user-specified value. At this point 
the initiator holds a structure with the relevant addressing and synchronization 
information (cf.~\Cref{fig:channel-initiator}), and the channel is 
ready to use.

\begin{table}
    \centering
    \begin{tabularx}{\columnwidth}{lX}
    \toprule
    \textbf{API} & \textbf{Purpose} \\
    \midrule
    \texttt{ramc\_put} & Put to target (blocking)\\
    \texttt{ramc\_put\_nb} & Put to target (non-blocking)\\
    \texttt{ramc\_await\_all\_puts} & Wait until expected number of PUTs complete (blocking)\\
    \texttt{ramc\_get} & Get from target (blocking)\\
    \texttt{ramc\_get\_nb} & Get from target (non-blocking)\\
    \texttt{ramc\_await\_all\_gets} & Wait until expected number of GETs complete (blocking)\\
    \bottomrule    
    \end{tabularx}
    \vspace{1pt}
    \caption{RAMC Communication Operations (called only by initiator).}
    \label{tab:communication-api}
\end{table}

Data movement operations are listed in~\Cref{tab:communication-api}. These come in blocking and non-blocking flavors. For 
example, \texttt{ramc\_put} blocks on a single increment of the local 
\texttt{FI\_WRITE} endpoint counter. That is, the call only returns when the 
initiator receives an acknowledgment from the target NIC that the data 
has been received, so the source buffer can be reused. RAMC uses the 
CXI default \texttt{FI\_TRANSMIT\_COMPLETE} completion semantics. In contrast, 
\texttt{ramc\_put\_nb} is non-blocking: 
A call to this function issues a \texttt{fi\_write} operation and increments 
a local \texttt{expected\_write\_counter\_value} to record that a write was 
issued. Later, a call to \texttt{ramc\_await\_all\_puts} blocks on the local 
\texttt{FI\_WRITE} counter reaching this expected value. Both versions of \texttt{put} utilize 
\texttt{fi\_inject\_write} for messages that satisfy the maximum inject size threshold (192B on Slingshot), and 
standard \texttt{fi\_write} otherwise.

%% file: example.tex
\section{Example}
\label{sec:example}

In this section we walk through a simple example illustrating how RAMC decouples initiator and target. 
To simplify exposition, error checking and some setup code has been removed.

\begin{lstlisting}[caption=RAMC put example, label={lst:ramc-put}, language=C]
unsigned char init_buf[BUFFER_SIZE];
unsigned char tgt_buf[BUFFER_SIZE];
struct ramc_target_win_info_s my_target_info;
struct ramc_init_win_info_s peer_info;

ramc_init();

if (1 == my_rank) {
    ramc_tgt_create_window(tgt_buf, BUFFER_SIZE, TAG, &my_target_info, INIT_STATUS_VAL);
    ramc_tgt_post_window(&my_target_info);
    ramc_tgt_activate_bb();
    ramc_tgt_await_bb_reads(1);
    ramc_tgt_deactivate_bb();
} else {
    do {
        ret = ramc_init_check_bb_status(peer_rank, tag);
    } while (ret != RAMC_SUCCESS);
    ramc_init_get_bb_posting(peer_rank, &peer_info, INIT_STATUS_VAL);
}

if (0 == my_rank) {
    ramc_init_increment_status(&peer_info, 1);
    do {
        ret = ramc_init_check_win_status(&peer_info);
    } while (ret != RAMC_SUCCESS);
    ramc_put(buffer, BUFFER_SIZE, 0);
    ramc_init_increment_status(&peer_info, 1);
} else {
    <WORK>
    ramc_tgt_increment_win_status(&,y_target_info, 1);
    <WORK>
    ramc_tgt_await_win_ops(&my_target_info, 1);
    ramc_tgt_increment_win_status(&my_target_info, 1);
}

ramc_barrier_binary();
if (1 == my_rank) {
    ramc_target_destroy_window(&my_target_info);
}
ramc_finalize();
\end{lstlisting}

\Cref{lst:ramc-put} Illustrates an initiator (rank 0) putting to a target (rank 1). 
Lines 1-4 are the primary data structures required for this example. The \texttt{init\_buf} contains 
the source data transmitted by the initiator, and the \texttt{tgt\_buf} is the 
target for this data.  The instance of \texttt{ramc\_target\_win\_info\_s} 
declared on line 3 is used by the target to 
store information regarding \texttt{tgt\_buf}, such as its size, the tag 
used for channel creation, libfabric addressing information for the 
target buffer and its status, etc. Likewise, the 
\texttt{ramc\_init\_win\_info\_s} structure (line 4) holds addressing 
information for accessing the target buffer and tracking its status.

RAMC is initialized on line 6; called by all processes, \texttt{ramc\_init} does basic libfabric setup (select the CXI provider, 
exchange endpoint addressing information, etc.), creates the BB for each process, and exchanges BB data and status addressing 
information so that every process can access the BB of every other process. 
This is a blocking operation.

On lines 9-13, the target creates and makes available information regarding a 
target window. On line 9, the target creates 
the window using \texttt{tgt\_buf}, specifying its size (in bytes), 
the tag to be 
used during channel creation, the structure in which to store target 
window information, and the initial target status value. This function caches 
the address of the buffer, its size, the tag, and creates a status buffer 
(initialized to read only). The function also registers and obtains addressing 
information for both the status and the data buffer via libfabric, and 
creates a memory region counter associated with \texttt{tgt\_buf}. The 
target then posts the addressing information for 
\texttt{tgt\_buf} and its status to the BB on line 10, changes the BB status to 
active (line 11), waits (blocking) for a 
single read of the BB information (line 12), and deactivates the BB after the 
read has occurred (line 13). 
On lines 15-18, the initiator polls the target's BB status until the target activates it (at which point the 
test returns \texttt{RAMC\_SUCESS}), and reads the posted addressing information into \texttt{peer\_info} (line 18). Checking the BB status/matching the tag 
must wait for data to be returned from the target, but is otherwise a non-blocking operation, offering an opportunity for the initiator to 
do other work if available.

The data transfer occurs over lines 21-34. On the initiator side (rank 0), 
the initiator first increments its local status value by one to indicate it 
expecting the target buffer to enter a write-enabled state (line 22; see~\Cref{sec:design-channels} for a description of how the examples in this 
article utilize the target and initiator status values). On lines 23-25, the 
initiator queries the status of the target buffer until it enters an 
\texttt{OK\_TO\_WRITE} phase as indicated by the status value of the target 
being equal to that of the initiator. Note that while the initiator must 
wait to receive data regarding the target's status, the target status check 
is otherwise non-blocking; at this point an application thus has 
an opportunity to perform other work, or to 
iterate through multiple target buffers looking for any that are ready to receive. 

When the status of the target is equal to that of the initiator (so the 
target is in a \texttt{OK\_TO\_WRITE} state), the initiator 
performs the put (line 26) and increments its local status value (line 27), indicating the initiator now expects the 
target buffer to enter a read only state. Note that the initiator is not limited to performing one communication operation; 
multiple operations can be issued before changing state.

On the target side, we note (line 29) that the target may be engaged in work when the initiator initially checks 
the target status (\Cref{fig:write-example}, steps 1 and 2). When this hypothetical work is complete, the target 
increments the status value of the target window (line 30). Since the write operation performed by the initiator requires no 
target involvement, this offers another opportunity to do additional work, as noted on line 31. When work is complete, 
the initiator waits for the single write to complete (line 32), and then updates the status of the target window to 
indicate it is \texttt{OK\_TO\_READ} (and hence will accept no more writes).

Lines 36-40 tear down what was set up for the example. 
The barrier (line 36) is a helper function implemented using 
a binary tree, and is not optimized. The target destroys the target window on line 38, which de-registers memory registered with libfabric and sets 
the status to a special 'destroyed' mode; if an initiator subsequently checks the status of this target buffer, the read status 
thereby informs the initiator the buffer no longer exists. The call to \texttt{ramc\_finalize} (line 40) 
cleans up any remaining RAMC and libfabric resources.

%% file: evaluation.tex
\section{Evaluation}
\label{sec:evaluation}

Fundamental RAMC capabilities and performance were assessed on Sandia 
National Laboratories' Eldorado supercomputer. Eldorado is an HPE Cray 
EX4000 system with 384 AMD MI-300A nodes. Each node has two sockets 
and four Slingshot 200 Gb/s NICs. To reduce system noise, a subset of cores 
on each node is reserved for the OS and other background processes 
\cite{leon_breaking_2025}, yielding 84 cores (out of 96) available for 
user processes. For tests involving MPI, we used Cray MPICH 9.0.1, the 
default on Eldorado. The system offers two versions of libfabric: a vendor-specific 
branch of libfabric 1.15.2 (known as Cray libfabric 2.1.3), and the open-source 
libfabric 2.3.1. The former has been superseded by the latter, but both are 
in use on Eldorado. Both libfabric versions have their own versions of the CXI 
libfabric provider.

\begin{figure}
\centering  
\includegraphics[width=\columnwidth]{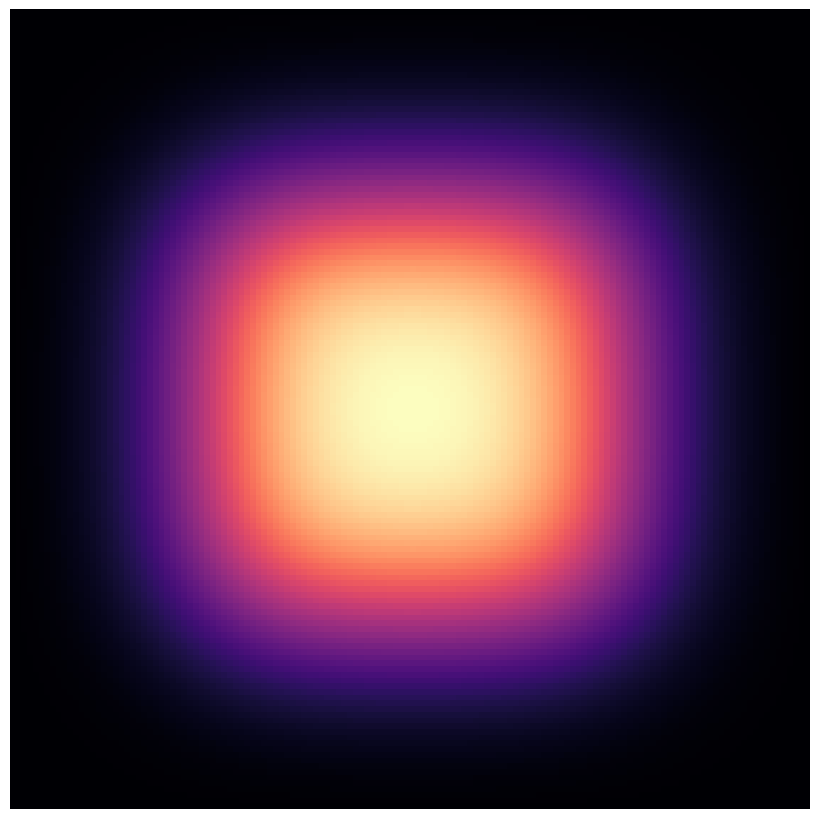}
\caption{Iteration 1000 from a heat diffusion code using RAMC for 5-point stencil communication. Data is for 19600 processes accross 250 nodes.}
\label{fig:hd_frame}
\end{figure}  

As noted above, RAMC is initially implemented with passive targets, 
which can generate significant network traffic while initiators poll 
targets for permission to write. As a test of RAMC's ability to scale, we implemented a standard heat diffusion code using RAMC 
for 5-point stencil communication. After a process has updated its internal 
temperature, it increments its local status value (thereby telling neighbors they can write their temperatures to the process' target window), and queries its neighbors -- north, east, south, west -- to 
determine whether a direction is ready to receive the updated value (through  
calls to \texttt{ramc\_init\_check\_win\_status}). When a neighbor is 
\texttt{OK\_TO\_WRITE}, the process writes its data to that process. When all 
neighbors have been serviced, the process waits on its own target counter 
(expecting four writes, one from each neighbor). When all expected writes 
have occurred, the process updates its temperature, and the cycle repeats. 
RAMC experienced no issues scaling this heat diffusion code to 19600 processes 
across 250 nodes (79 processes per node maximum;~\Cref{fig:hd_frame}).

\begin{figure}
\centering  
\includegraphics[width=\columnwidth]{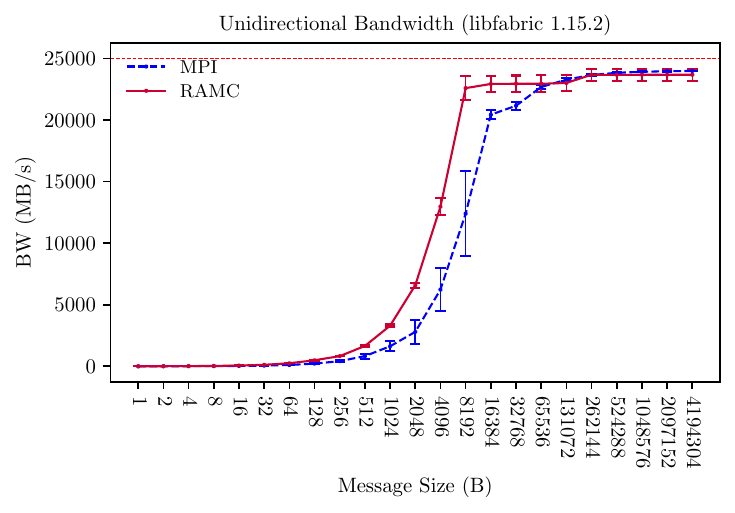}
\caption{Unidirectional bandwidth under libfabric 1.15.2 for OMB MPI P2P (Cray MPICH) and RAMC using counters. Dashed red line is theoretical maximum.}
\label{fig:bw-1.15.2}
\end{figure}  

\begin{figure}
\centering  
\includegraphics[width=\columnwidth]{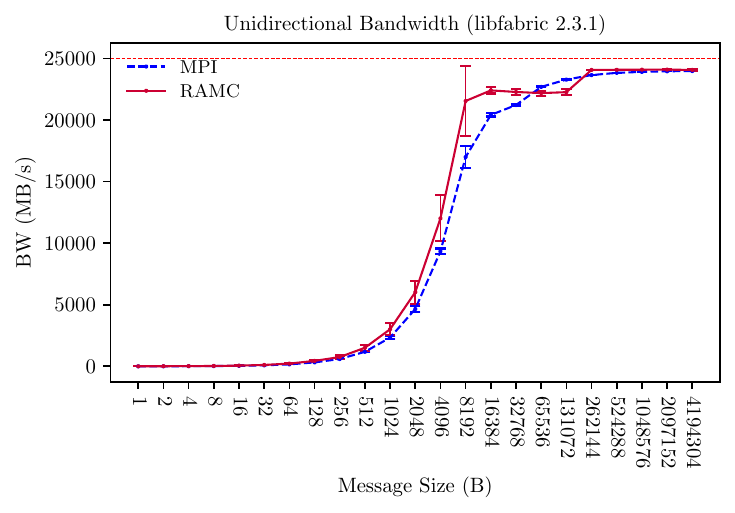}
\caption{Unidirectional bandwidth under libfabric 2.3.1 for OMB MPI P2P (Cray MPICH) and RAMC using counters. Dashed red line is theoretical maximum.}
\label{fig:bw-2.3.1}
\end{figure}  

HPE's Cray MPICH is highly optimized for Slingshot, and Portals-based 
NICs ultimately use RMA operations for data movement~\cite{portals_43}. 
With this in mind, we wrote unidirectional bandwidth and ping-pong latency tests for RAMC that 
mirror those of the well-known OMB micro-benchmarks, for 
purposes of comparison with Cray MPICH MPI two-sided results~\cite{OMB}. We 
opted to compare with OMB two-sided bandwidth and latency rather than OSU one-sided 
benchmarks because the latter target performance characteristics of less immediate interest. For 
example, the OMB put latency benchmark measures the `local' latency of an \texttt{MPI\_PUT} call, i.e., 
the time lapsed before the source buffer can be reused, rather than the latency of message passing. 
All results are averages of ten runs with error bars indicating standard deviation.

\Cref{fig:bw-1.15.2} compares RAMC unidirectional bandwidth with OMB results, under 
libfabric 1.15.2. RAMC consistently outperforms MPI, showing increases in mean bandwidth 
ranging from approximately 100-130\% for messages in the range of 1 to 4KiB, at which point 
the gains begin to recede to parity by 32KiB. 

The gap is less prominent under libfabric 2.3.1, as shown in \Cref{fig:bw-2.3.1}. In this case, 
RAMC bandwith is approximately 30-45\% greater than MPI for messages in the range of 1-8KiB, and 
MPI holds a slight lead (2-4\%) for 64KiB and 128KiB messages. This change is due to both an 
apparent increase in mean bandwidth for MPI and a slight decrease in the same for RAMC with the switch from 
libfabric 1.15.2 to 2.3.1. For example, MPI shows increases in mean bandwidth in the range of 32-66\% for 
1B-8KiB messages, while RAMC bandwidths drop by up to 11\% over the same range. 

\begin{figure}
\centering  
\includegraphics[width=\columnwidth]{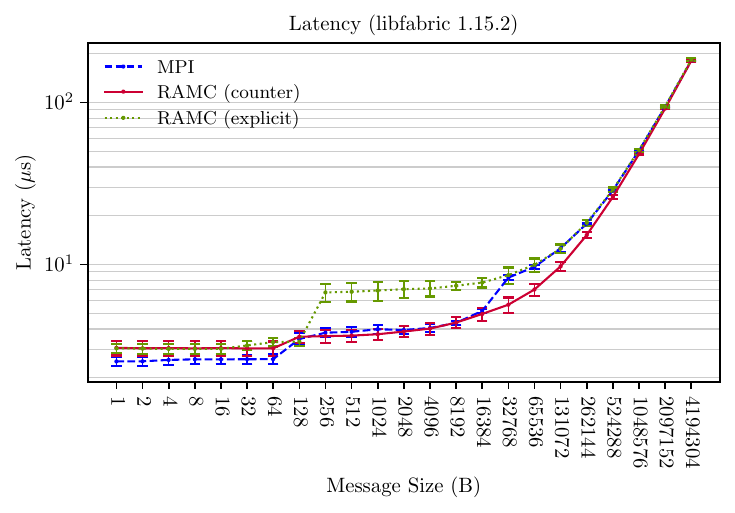}
\caption{Ping-pong latencies under libfabric 1.15.2 for OMB MPI P2P (Cray MPICH), RAMC using counters, and RAMC 
using explicit notification.}
\label{fig:latency-1.15.2}
\end{figure}

\begin{figure}
\centering  
\includegraphics[width=\columnwidth]{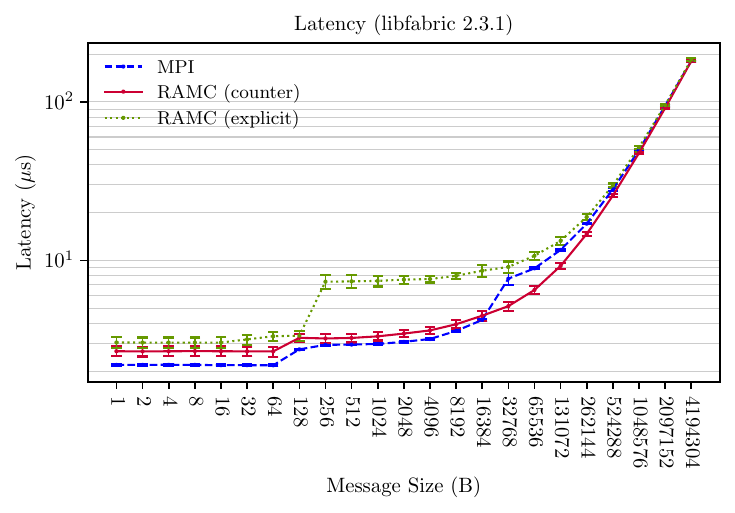}
\caption{Ping-pong latencies under libfabric 2.3.1 for OMB MPI P2P (Cray MPICH), RAMC using counters, and RAMC 
using explicit notification.}
\label{fig:latency-2.3.1}
\end{figure}

A benefit of Slingshot is lightweight target-side RMA completion notification via memory region counters. A 
more traditional approach uses explicit completion notification in the form of a second RMA operation following the first; this 
followup operation ensures completion of the first through ordering constraints, and lets the target know the preceding operation 
has occurred. To better understand the impact of Slingshot MR counters on latencies, we implemented an explicit notification 
version of RAMC by adding a notification buffer to each target window; after issuing a write to the data buffer, the initiator 
performs an atomic increment (through libfabric atomics) on this notification buffer. Instead of testing on a MR counter, the 
target checks this notification buffer for the expected number of increments. Note this explicit notification mechanism is simple 
to build by creating an additional RAMC channel for notifications.

\Cref{fig:latency-1.15.2} shows ping-pong latency results using libfabric 1.15.2 for three cases for MPI, standard RAMC (i.e., using MR counters for completion), and 
RAMC with explicit notification. Explicit notification latencies are nearly indiscernible from those of standard RAMC for 1-128B messages, at 
which point the latencies of the former jump, showing an 86\% increase at 256B. The gap between explicit and standard decreases as 
message sizes grow (e.g., 52\% at 32KiB), until there is little difference by 1MiB. Note the location of the increase in latencies for 
explicit notification corresponds to the maximum inject size (192B) for Slingshot.

For small message sizes (1-64B), Cray's MPI shows a distinct advantage in mean latencies, with RAMC lagging by 430-540 ns. By the 
inject threshold (192B), this gap is eliminated. RAMC subsequently shows reductions in latencies (relative to MPI) of 
up to 32\% for 16-512KiB messages, where 16KiB is the eager-rendezvous threshold.

\Cref{fig:latency-2.3.1} shows latency results under libfabric 2.3.1. In this case, explicit notification exhibits an approximate 
latency penalty of 13-24\% compared to standard RAMC for small messages (1-64KiB), and the jump at the inject limit surges to 
a 128\% increase. Cray's MPI continues to exhibit lower latencies than standard RAMC for small messages; these are of the same magnitude 
as for 1.15.2: 480-500 ns. In contrast, under 2.3.1 this trend continues into medium-sized messages, with Cray's MPI averaging approximately 
339 ns faster than standard RAMC. As with 1.15.2, crossing the eager-rendezvous threshold incurs penalties for MPI, and RAMC shows 
decreases in latencies of up to 32\% for 16-512KiB messages.

Comparing MPI and RAMC latencies across libfabric 1.15.2 and 2.3.1, we observe Cray's MPI shows decreases in latencies for 1-16KiB messages of 
13-26\%, while RAMC shows decreases of 10-13\%. These decreases reflect additional performance tuning within the CXI provider, but also 
reveal these optimizations have a greater impact on Cray's MPI than RAMC. For this work, we considered several opportunities for addressing this 
disparity, including (i) carefully limiting requested provider capabilities to the minimum required for RAMC; (ii) confirming that Cray MPI CXI 
defaults correspond to those used by RAMC; (iii) explicitly disabling receiver-side message processing ordering constraints that appear to be 
enabled by default (as indicated by the structure returned by \texttt{fi\_getinfo}), and (iv) using an aliased endpoint configured to use a 
low-latency traffic class (\texttt{FI\_TC\_LOW\_LATENCY}) for transmitting small messages: None of these strategies showed any improvement to RAMC latencies, 
and in some cases increased them (case (iv)). 

Finally, we also 
investigated whether RAMC latencies were the result of a reliance on MR counters. Specifically, Slingshot provides two types of memory 
regions: `standard' and `optimized'. The latter (identified by user-specified memory keys in the range of 0-99) allow initiators to use smaller 
headers, reducing latencies for small writes, and are intended for applications that use a small number of large MRs. However, they are not intended 
for use with MR counters (i.e., \texttt{FI\_RMA\_EVENT} should not be an enabled provider capability). To investigate the potential impact of 
optimized MRs on latencies, we gutted all use of MR counters from RAMC (for both BB and channels), replacing all counter completion with 
explicit notification and disabling the \texttt{FI\_RMA\_EVENT} capability. Unfortunately, using optimized MRs with explicit notification 
incurred increases in latencies of up to 60\% for small messages (1-128B), and of up to 160\% for medium messages (256B-16KiB), compared to 
standard RAMC. Consequently, based on this initial exploration, optimized MRs do not appear to be responsible for Cray MPI's reduced latencies. 

%% file: related.tex
\section{Related Work}
\label{sec:related-work}

RAMC provides an explicit one-sided communication model, and there has been significant prior work 
in this area. A prominent example is the MPI RMA API, which provides explicit one-sided communication, and has been under development 
for decades. Introduced in MPI 2.0 with an initial set of functionality, the RMA API was expanded in 
MPI 3.0 to include passive target synchronization~\cite{mpi20,mpi30}, and MPI RMA has seen a nuber of efforts in improving 
performance, such as in multithreaded use cases~\cite{hjelm2018improving}. In contrast with RAMC, however, 
MPI RMA treats windows as monolithic (i.e., all processes participate in window creation and synchronization), and 
lacks a standardized RMA-with-notification semantic as provided by Slingshot's MR counters. Note, however, there is 
currently an RMA-with-notification proposal under consideration for inclusion in an upcoming revision of the 
MPI specification~\cite{belli_notified_2015}. From this perspective, RAMC can be viewed as one approach towards 
implementing MPI RMA-with-notification, although exploring this possibility is left for future work.

OpenSHMEM is another major explicit one-sided communication model targeting high performance 
computing~\cite{Poole2011}. It has also seen major interest including for GPU-based communication with the 
NVSHMEM implementation~\cite{hsu2020initial}. Like RAMC (and unlike MPI) OpenSHMEM and its variants (NVSHMEM, CraySHMEM) 
include signaled RMA operations designed to avoid an explicit follow-up notification message, although how this 
is implemented will differ across network hardware (e.g., an alternative to counters is allowing incoming RMA 
operations to force events to appear on a completion queue)~\cite{jose2014designing}. In contrast to RAMC, OpenSHMEM still 
treats windows as monolithic constructs, shared across all processes in the group (e.g., as a symmetric heap).

A core aspect of RAMC is that it provides genuinely persistent communication channels. MPI also includes 
API calls that invoke persistence, although there are important differences in comparison to RAMC. 
Despite the name, MPI persistent point-to-point communication (\texttt{MPI\_Send\_init, MPI\_Recv\_init, MPI\_Start}) 
involves a distinct tag matching step on each transmission (e.g., one can match a call to \texttt{MPI\_Start} with 
an target-side \texttt{MPI\_Irecv}); the persistence is really a form of information caching rather than a fixed target 
buffer as with RAMC. In contrast, MPI Partitioned Communication, introduced in MPI 4.0~\cite{mpi40}, aims to 
leverage the benefits of true persistence by creating a permanent match across a channel that can 
be reused multiple times~\cite{grant2019finepoints,gillis2023quantifying,temuccin2024design}. In contrast to 
RAMC, partitioned communication is more restricted in that it utilizes datatypes and counts rather than bytes, and offsets are constrained 
to partition boundaries. Moreover, under the current MPI specification, an implementation of 
partitioned communicaton over RMA will not be specification compliant~\cite{dosanjh2021implementation}.

Finally, we note RAMC is closely related to research on triggered communication operations (i.e., deferred work 
queues). Specifically, besides providing lightweight RMA completion notification on a target, Portals-style 
MR counters provide a mechanism for offloading communication progress to the NIC. Communication operations can be 
associated with an MR counter and given a threshold; when the counter is incremented, all operations whose threshold is 
less-than-or-equal-to the current counter value are automatically executed by the NIC, independent of the host CPU. Originally 
proposed to facilitate the offloading of collectives and rendezvous messaging, this mechanism has recently been 
adopted to allow GPU-triggered communication as well~\cite{hemmert_using_2010,barrett_using_2011,namashivayam_exploring_2022,bridges2026codesignevaluationcpufreempi}. While triggered operations are not part of the 
current RAMC API, they will be considered in future revisions.

%% file: challenges.tex
\section{Challenges and Future Work}
\label{sec:challenges}

In this work we've presented RAMC, a communication library that leverages 
the hardware capabilities of Cray HPE's Slingshot interface to provide efficient and 
easy-to-use unidirectional RMA-based channels. In contrast to existing user-level 
RMA APIs, RAMC does not require monolithic window creation or synchronization, enabling 
point-to-point style persistent communication semantics as well as opportunities for earlybird 
communication.

An obvious limitation of RAMC is that it relies on hardware capabilities of the HPE Cray Slingshot 
network. On the one hand, this is deliberate. RAMC is not proposed as a general-purpose user-level 
communication API, but rather as an exploration of capabilities 
provided by the Slingshot network. If these capabilities turn out to be especially useful or 
performant, there is reason to advocate for wider adoption. On the other, there is nothing in RAMC that precludes implementation 
using other underlying mechanisms, such as substituting explicit notifications for Slingshot MR counters. 

A second potential point of concern is that the current implementation is \emph{passive target} insofar 
as an initiator queries a target's status until that status indicates the initiator may perform its 
operations. As noted in~\Cref{sec:evaluation}, this situation has the potential to generate significant 
traffic (cf.~\cite{jiang_efficient_2004}). An alternative is an \emph{active target} strategy, where 
an initiator instead checks for a `ready-to-operate' signal from a target. This approach would 
reduce synchronization network traffic, and remove a round trip communication before an initiator 
issues a communication operation.

A third challenge for RAMC is imposed by libfabric (and Slingshot). Because target-side 
MR counters are associated with specific memory regions, a target can distinguish between 
operations on different MRs. In contrast, the endpoint completion counters used by 
initiators do not distinguish between targets, because they count \emph{all} operations of 
a given type issued on that endpoint. Consequently, for example, if an initiator 
issues writes to two buffers on the same target process, it cannot test for 
completion of one or the other; it must wait for both to complete to conclude that 
either has. Bringing the granularity of initiator-side completion counters 
into alignment with those of the target would enable more fine-grained coordination between 
initiators and targets.

Finally, MR counters in Slingshot are not limited to signaling completion but are also 
part of a strategy for offloading progress to the NIC. Specifically, communication operations 
can be registered with a Slingshot NIC but deferred until an associated counter 
reaches a specified threshold. When this threshold is reached, the operation is 
triggered, with no host involvement. Originally intended for the offloading of 
collective and rendezvous operations~\cite{hemmert_using_2010,underwood_enabling_2011,barrett_using_2011}, 
this capability has more recently been leveraged to enable GPUs to initiate 
communication without host involvement~\cite{namashivayam_exploring_2022,bridges2026codesignevaluationcpufreempi}. 
Future work on RAMC will explore further leveraging the completion counters 
intrinsic to RAMC's operation to offload communication progress.